\documentclass[epj,final,nopacs]{tjb}
\usepackage{graphics,graphicx,amsfonts,amsmath,pstricks}

\def\>{\big>}
\def\<{\big<}
\def\|{\big\vert}
\def\£{\big\Vert}
\def\>{\big>}
\def\<{\big<}
\def\){\big)}
\def\({\big(}

\def\etal{\textit{et~al}}

\def\rf#1{(\ref{#1})}

\def\etab{\end{tabular}}  
  
\def\bit{\begin{itemize}}
\def\eit{\end{itemize}}

\def\bml{\begin{multline}}
\def\eml{\end{multline}}
\def\be{\begin{equation}} 
\def\ee{\end{equation}}  
\def\bea{\begin{eqnarray}}  
\def\eea{\end{eqnarray}} 
\def\bmu{\begin{multline}}
\def\emu{\end{multline}}
\def\bal{\begin{array}{l}} 	
\def\eal{\end{array}}

\def\ot{\times}

\def\thalf{\tfrac{1}{2}}

\def\MeV{\textrm{ MeV}}

\def\x{\ve{x}}

\renewcommand{\L}{\Lambda}

\def\S{\Sigma}
\def\P{\Pi}

\def\letterS{S}
\def\letterP{P}
\def\letterD{D}
\def\letterF{F}
\def\greek#1{\def\letter{#1}
\ifx\letter\letterS\Sigma
\else
	\ifx\letter\letterP\Pi
	\else
		\ifx\letter\letterD\Delta
		\else
			\ifx\letter\letterF\Phi
			\else XXX
			\fi
		\fi
	\fi
\fi
}

\def\q{\bar q}

\def\c{\bar c}
\def\nn{n\bar n}

\def\uu{u\bar u}
\def\ud{u\bar d}

\def\uc{u\bar c}

\def\dd{d\bar d}

\def\dc{d\bar c}

\def\dd{d\bar d}

\def\cc{c\bar c}

\def\dc{d\bar c}
\def\sc{s\bar c}
\def\cc{c\bar c}

\def\MeV{\textrm{ MeV}}

\renewcommand{\u}[1]{\rm{#1}}
\def\cn#1#2#3{^{#1}{\u{#2}}_{#3}}     
\def\an{\cn} 
\def\sl#1#2{^{#1}{\u{#2}}}    
    
\def\Jp{J/\psi}

\def\epem{e^+e^-}

\def\Ph{P_c(4450)^+ }
\def\Pl{P_c(4380)^+ }
\def\P{P_c}
\def\L{\Lambda}
\def\Lb{\Lambda_b^0}
\def\Lc{\Lambda_c}
\def\Sc{\Sigma_c}
\def\Db{\bar D}
\def\Jp{J\!/\!\psi}
\def\D{\Delta}
\def\*{^{(*)}}
\def\SDl{\Sc^* \Db}
\def\SDh{\Sc\Db^*}
\def\SD{\Sc\*\Db\*}
\def\LD{\Lc^*\Db}
\def\JpN{\Jp N^*}
\def\chip{\chi_{c1}p}
\def\t{$\checkmark$}
\def\x{$\times$}

\begin{document}
\sloppy 
\title{Phenomenology of $P_c(4380)^+$, $P_c(4450)^+$ and related states}
\author{T. J. Burns}
\institute{Department of Physics, Swansea University, Singleton Park, Swansea, SA2 8PP, UK\\
{\tt t.burns@swansea.ac.uk}}
\date{~}
\abstract{The $P_c(4380)^+$ and $P_c(4450)^+$ states recently discovered at LHCb have masses close to several relevant thresholds, which suggests they can be described in terms of meson-baryon degrees of freedom. This article explores the phenomenology of these states, and their possible partners, from this point of view. Competing models can be distinguished by the masses of the neutral partners which have yet to be observed, and the existence or otherwise of further partners with different isospin, spin, and parity. Future experimental studies in different decay channels can also discriminate among models, using selection rules and algebraic relations among decays. Among the several possible meson-baryon pairs which could be important, one implies that the states are mixtures of isospins 1/2 and 3/2, with characteristic signatures in production and decay. A previous experimental study of a Cabibbo-suppressed decay showed no evidence for the states, and further analysis is required to establish the significance of this non-observation. Several intriguing similarities suggest that $P_c(4450)^+$ is related to the $X(3872)$ meson.} 
\maketitle
\section{Introduction}
\label{introduction}
The recent discovery at LHCb~\cite{Aaij:2015tga} of states decaying to $\Jp p$ has provoked considerable excitement. The $\Pl$ and $\Ph$ states are observed in $\Lb\to \Jp p K^-$. For the heavier state in particular, there is a dramatic peak in the $\Jp p$ invariant mass spectrum and clear evidence of phase motion.

Already there are several proposed interpretations of these states in the literature, some of which are inspired by their proximity to meson-baryon thresholds. The most prosaic option is that they are kinematic effects due to rescattering among different channels~\cite{Guo2015,Liu2015,Mikhasenko2015,Meissner2015}. Alternatively, they could be bound states (or resonances) formed from open-charm baryon and meson constituents~\cite{arxiv-1507.03717,arxiv-1507.03704,He2015,arxiv-1507.04249,Xiao2015}, a possibility which had been anticipated in several models~\cite{Wu2010,Yang2011,Xiao:2013yca,Karliner2015}, or baryocharmonia~\cite{Kubarovsky:2015aaa}. The compact pentaquark interpretation ignores the role of thresholds and describes the states in terms of quark, diquark or triquark degrees of freedom~\cite{Yuan:2012wz,Maiani:2015vwa,Anisovich2015a,Ghosh2015,Lebed:2015tna,Li2015,Wang2015b,Wang2015c}. Other possibilities have also been discussed~\cite{Mironov2015,Scoccola2015}.

This article explores the implications of meson-baryon degrees of freedom for $\Pl$, $\Ph$ and their possible partner states.
The experimental properties of the states are first compared with naive theoretical arguments (Sec.~\ref{ass}). The various possible meson-baryon thresholds which could play a role are introduced, and it is noted that none of the possibilities fits entirely with naive expectations for S-wave couplings (Sec.~\ref{degrees}). Models are introduced and their predictions are confronted with experimental data (Sec.~\ref{models}). The observed charged states should be accompanied by neutral partners, and possibly additional partners with different $J^P$ or isospin (Sec.~\ref{partners}). One of the possible meson-baryon combinations is shown to imply isospin violation, due to a mass gap separating different charge channels (Sec.~\ref{isospin}). Many decays other than the observed $\Jp p$ are possible, and it is argued that characteristic patterns among these due to isospin and heavy-quark spin can discriminate among models (Sec.~\ref{decays}). Using a simple model relations among different decay modes are obtained, and these suggest that isospin-violating $\Jp \D$ and $\eta_c\D$ decays could be large (Sec.~\ref{relations}). The states, and their missing partners, may also be observed in other channels, including Cabibbo-suppressed decays where data are already available (Sec.~\ref{production}). Finally, it will be shown that there are several intriguing parallels between $\Ph$ and $X(3872)$, suggesting that they may be related (Sec.~\ref{molecule}).

\section{Spin-parity assignments}
\label{ass}

\begin{table}
\caption{The masses and widths (MeV) of the $P_c$ states, their possible $J^P$  assignments, and the energies (MeV) of nearby thresholds. Here $\Sc$, $\Sc^*$ and $\Lc^*$ are $\Sc(2455)$, $\Sc(2520)$ and $\Lc(2595)$, respectively.}
\label{props}       
\begin{tabular}{p{2.5cm}p{2.5cm}p{2.5cm}}
\hline\noalign{\smallskip}
& $\Pl$& $\Ph$  \\
\noalign{\smallskip}\hline\noalign{\smallskip}
Mass &$ 4380\pm 8\pm $29 & $4449.8\pm 1.7\pm 2.5 $\\
Width & $205\pm 18\pm 86 $&$ 35\pm 5\pm 19$ \\
\noalign{\smallskip}\hline\noalign{\smallskip}
Assignment~1 &$3/2^-$&$5/2^+$\\
Assignment~2 &$3/2^+$&$5/2^-$\\
Assignment~3 &$5/2^+$&$3/2^-$\\
\noalign{\smallskip}\hline\noalign{\smallskip}
$\Sc^{*+}\Db^0	$&$4382.3\pm 2.4$\\
$\chip$			&&$4448.93\pm 0.07$\\
$\Lc^{*+}\Db^0$	&&$4457.09\pm 0.35$\\
$\Sc^+\Db^{*0}$	&&$4459.9\pm 0.5$\\
$\Sc^+\Db^0\pi^0$&&$4452.7\pm 0.5$\\
\noalign{\smallskip}\hline
\end{tabular}
\end{table}

The properties~\cite{Aaij:2015tga} of the new states are summarised in Table~\ref{props}.  For convenience, $\Pl$ and $\Ph$ will be referred to throughout this paper as $\P$ and $\P^*$, respectively. The $J^P$ assignments  are not yet determined definitively, but the best fit to data (Assignment~1 in the table) is obtained with $\P$ and $\P^*$ as $3/2^-$ and $5/2^+$ states, respectively. The $(3/2^+,5/2^-$) combination (Assignment~2) and ($5/2^+,3/2^-$) combination (Assignment~3) are also possible. The discussion in this paper will concentrate on these three assignments, as they are preferred experimentally, although other possibilities cannot yet be ruled out.

The proximity of nearby meson-baryon thresholds plays an important role in several models for the $\P\*$ states. A summary of the relevant threshold energies is given in Table~\ref{props}; in that table, and elsewhere in this paper, $\Sc(2455)$, $\Sc(2520)$ and $\Lc(2595)$ will be referred to as $\Sc$, $\Sc^*$ and $\Lc^*$, respectively.  The $\P^*$ peak overlaps with the $\chip$  threshold~\cite{Guo2015}, and is somewhat below $\LD$~\cite{Aaij:2015tga} and $\Sc\Db^*$~\cite{He2015}, while $\P$ is somewhat below $\Sc^*\Db$~\cite{He2015}. The proximity of $\P^*$ to the three-body $\Sc^+\Db^0\pi^0$ channel has not been noted in the literature: this follows automatically from its proximity to $\Sc^+\Db^{*0}$, since $\Db^{*0}$ is just above $\Db^{0}\pi^0$ threshold.

Their $\Jp p$ decays suggest that, regardless of the internal dynamics, the states have intrinsic quark content $uudc\c$. As a point of reference, Table~\ref{partials} summarises the different spin and isospin channels available to this five quark system, classified according to the two possible meson-baryon decompositions: open-charm  $(udc)(u\c)/(uuc)(d\c)$ and closed-charm $(uud)(c\c)$. The entries S, P, D or F denote the allowed partial waves in which the given meson-baryon pair couples to the appropriate $J^P$. The three experimental $J^P$ assignments for $\P$ and $\P^*$ are identified at the top of the table.

\begin{table}
\caption{The isospin and $J^P$ combinations accessible to different open- and closed-charm meson-baryon systems. The table entries are the allowed partial waves (up to $L=3$) for each channel.  The experimental $J^P$ assignments for $\P$ and $\P^*$ are indicated in the first three rows.}
\label{partials}       
\begin{tabular}{llllllll}
	\hline
	\noalign{\smallskip}                                                       &                & $\frac{1}{2}^-$ & $\frac{3}{2}^-$ & $\frac{5}{2}^-$ & $\frac{1}{2}^+$ & $\frac{3}{2}^+$ & $\frac{5}{2}^+$ \\
	\noalign{\smallskip}\hline\noalign{\smallskip}                             & Ass. 1         &                 & $\P$            &                 &                 &                 & $\P^*$          \\
	                                                                           & Ass. 2         &                 &                 & $\P^*$          &                 & $\P$            &  \\
	                                                                           & Ass. 3         &                 & $\P^*$          &                 &                 &                 & $\P$            \\
	\noalign{\smallskip}\hline\noalign{\smallskip}
$I=\frac{1}{2}$             & $\Lc \Db$      & S               & D               & D               & P               & P               & F               \\
	                                                                           & $\Lc \Db^*$    & SD              & SD              & D               & P               & PF              & PF              \\
	                                                                           & $\Lc(2595)\Db$ & P               & P               & F               & S               & D               & D               \\
	\noalign{\smallskip}\hline\noalign{\smallskip}
$I=\frac{1}{2},\frac{3}{2}$ & $\Sc \Db$      & S               & D               & D               & P               & P               & F               \\
	                                                                           & $\Sc^* \Db$    & D               & SD              & D               & P               & PF              & PF              \\
	                                                                           & $\Sc\Db^*$     & SD              & SD              & D               & P               & PF              & PF              \\
	                                                                           & $\Sc^* \Db^*$  & SD              & SD              & SD              & PF              & PF              & PF              \\
	\noalign{\smallskip}\hline\noalign{\smallskip}
$I=\frac{1}{2}$             & $\Jp N$        & SD              & SD              & D               & P               & PF              & PF              \\
	                                                                           & $\eta_c N$     & S               & D               & D               & P               & P               & F               \\
	                                                                           & $\chi_{c1} N$  & P               & PF              & PF              & SD              & SD              & D               \\
	                                                                           & $\chi_{c0} N$  & P               & P               & F               & S               & D               & D               \\
	                                                                           & $\Jp N(1440)$  & SD              & SD              & D               & P               & PF              & PF              \\
	                                                                           & $\Jp N(1520)$  & PF              & PF              & PF              & SD              & SD              & SD              \\
	\noalign{\smallskip}\hline\noalign{\smallskip}
$I=\frac{3}{2}$             & $\Jp \D$       & SD              & SD              & SD              & PF              & PF              & PF              \\
	                                                                           & $\eta_c\D$     & D               & SD              & D               & P               & PF              & PF              \\
	\noalign{\smallskip}\hline                                                 &
\end{tabular}
\end{table}

There are two naive arguments which suggest that of the different possible $J^P$ assignments, that preferred by experiment (Assignment~1) is also the most natural theoretically. Firstly, due to the centrifugal barrier suppressing decays in higher partial waves, it would be natural if $\P$, as the broader of the two states, decays in a lower partial wave than $\P^*$. (This assumes some similar underlying structure for the two states.) Referring to Table~\ref{partials}, the minimum allowed partial waves for the $\Jp p$ decays of $\P/\P^*$ in Assignments 1, 2 and 3  are respectively S/P-wave, P/D-wave and P/S-wave. On this basis Assignments 1 and 2 are consistent with expectations, while Assignment~3 is not.

Secondly, the intrinsic negative parity of the quark combination $uudc\c$ suggests that, regardless of the assumed internal dynamics, the negative parity states (involving no orbital excitations) will be lightest, while positive parity states (with one unit of orbital excitation) will be heavier. Only Assignment~1 is consistent with this ordering.

Neither of these arguments should be taken too seriously, particularly the second: the opposite parities of the states are a challenge in most models, regardless of the ordering.

\section{Meson-baryon degrees of freedom}
\label{degrees}

Ordinary hadrons ($qqq$ baryons and $q\q$ mesons) couple to two-body channels by quark-pair creation, which affects their masses and (above threshold) leads to strong decay. Quenched quark models which ignore these couplings work well for states far below threshold, but not because the two-body components and their effects are necessarily small. Large mass shifts due to two-body couplings can partly be absorbed into a redefinition of model parameters~\cite{Barnes:2007xu}, and the remaining induced mass splittings, while significant, leave several  quenched quark model results intact \cite{Burns:2011jv,Burns:2011fu,Burns:2012pc,Burns:2014qya}. Close to threshold, though, a two-body description is unavoidable: this is model-independent and is due to the associated small energy denominator. In the case of the near-threshold $X(3872)$ meson, for example, models which include both $\cc$ and $(c\q)(q\c)$ degrees of freedom, and allow for coupling between them, find that the wavefunction is dominated by $(c\q)(q\c)$ components \cite{Kalashnikova:2005ui,Li:2009ad,Kalashnikova:2010zz,Coito:2012vf,Takizawa:2012hy,Ortega:2012rs,Ferretti:2013faa}.

Similarly, the proximity of their masses to thresholds implies that the $\P\*$ states contain significant meson-baryon components in their wavefunctions, unless for some reason the coupling to those channels is suppressed. Models which ignore these components must not only accept as a coincidence the proximity of masses, but should also explain why the relevant couplings are small. Naively the coupling for a state which has the same valence quark content ($uud\cc$) as a two-body pair would be larger than the corresponding two-body couplings of ordinary $qqq$ baryons or $q\q$ mesons, which require quark-pair creation. 
 
Conventional wisdom is that two-body thresholds are most important in S-wave (although this is not necessarily true of kinematic models, discussed later). However, the experimental $J^P$ assignments do not allow for more than one of the $\P\*$ states to couple to the relevant threshold in S-wave. Referring to Table~\ref{partials}, for $\P^*$ the couplings to $\chip$ for the three assignments are respectively D-, P- and P-wave. Similarly, for $\P^*$ the couplings to $\LD$ threshold are D-, F- and P-wave. The $\P/\P^*$ couplings to their respective $\SDl/\SDh$ channels are, for the three assignments, S/P-wave, P/D-wave and P/S-wave. Assuming that lower partial wave couplings are more important, the relevance of $\chip$ seems most natural in Assignments~2 or 3, that of $\LD$ in Assignment~3, and that of $\SDl/\SDh$ in Assignments~1 or 3.

Meson-baryon components are also integral to the baryocharmonium interpretation~\cite{Kubarovsky:2015aaa} in which $\P$ and $\P^*$ are $\Jp N(1440)$ and $\Jp N(1520)$ composites, albeit with masses less closely correlated to corresponding thresholds. In this model Assignment~1 allows for both $\P\*$ states in S-wave.

In this paper it will be assumed that the $\P\*$ states are described in terms of meson-baryon degrees of freedom. The following scenarios will be considered: $\P^*$ as a $\chip$ state or $\LD$ state,  and $\P/\P^*$ as $\SDl/\SDh$ states or $\Jp N(1440)/\Jp N(1520)$ states. These will be referred to as the $\chip$, $\LD$, $\SD$ and $\JpN$ scenarios.

The term ``state'' is being used loosely here and, consistent with the general nature of much of the discussion below, does not presuppose any particular model. Most of the conclusions rely only on the assumed  spin and flavour degrees of freedom of the meson-baryon pair, not on the underlying dynamics that give rise to the states in the first place.

The scenarios outlined above will mostly be treated as distinct possibilities for the $\P\*$ states, although in reality there may be some interplay among the various degrees of freedom, particularly for $\P^*$ which has several nearby thresholds. Even if there is mixing among different wavefunction components, the conclusions can still be useful, as discussed in Sec.~\ref{molecule}.

Finally, note that in this simplified classification of competing, distinct scenarios, the $\chip$ and $\LD$ scenarios apply only to $\P^*$, whereas the $\SD$ and $\JpN$ scenarios, at least in principle, could accommodate the doublet of $\P\*$ states. 

\section{Models}
\label{models}
In this section competing models for the $\P\*$ states are introduced. Kinematic effects, discussed first, arise from $\chip$, $\LD$ and/or $\SD$ degrees of freedom, while models in which the states are genuine meson-baryon composites involve $\SD$ or $\JpN$ degrees of freedom.

Kinematic models appear to be a better match for $\P^*$ than $\P$. In the model of Guo~\etal.~\cite{Guo2015}, $\P^*$ is a kinematic effect associated with the $\chip$ threshold, assuming a P-wave coupling (Assignments 2 or 3), and arising from two possible mechanisms: a two-point singularity, where the decay $\Lb\to \chi_{c1}p K^-$ is followed by rescattering $\chi_{c1}p\to \Jp p$, and a triangle singularity due to $\Lb\to \chi_{c1}\Lambda(1890)$, $\L(1890)\to p K^-$, again followed by rescattering $\chi_{c1}p\to \Jp p$. In both mechanisms the $\chip$ combination feeds $\P^*$, so from the point of view of this paper $\P^*$ in this model is regarded as a $\chip$ state. A similar point of view is adopted in ref.~\cite{Meissner2015}.

Mikhasenko~\cite{Mikhasenko2015} also describes $\P^*$ as a kinematic effect due to a triangle singularity, but in this case the mechanism is $\Lb\to \Sc^+ D_s^{*-}$, $D_s^{*-}\to \Db^{*0} K^-$ followed by rescattering $\Sc^+\Db^{*0}\to \Jp p$. From the point of view of this paper, in this model $\P^*$ is regarded as a $\Sc\Db^*$ state.

Liu~\etal.~\cite{Liu2015} argue that peaks arising near the $\P^*$ mass can result from two-point and triangle singularities, and their calculations include a larger number of intermediaries $\chi_{cJ}p$, $\Lc\*\Db\*$ and $\Sc\*\Db\*$. If each of these intermediaries plays a comparable role then it is not possible to classify the $\P\*$ states uniquely according to the distinct scenarios outlined above. In particular, the $\chi_{c1}p$ and $\LD$ channels, through triangle singularities, appear to play a comparable role in the $\P^*$ peak (see their Fig.~5).

A different class of models regards $\P/\P^*$ as $\SDl/\SDh$ composites, with a binding potential due to meson exchange. A challenge common to all such models is that, due to their opposite parities, it is not possible to match both $\P\*$ states to S-wave thresholds. The least problematic interpretations (as remarked above) are Assignments 1 and 3, which have $\P/\P^*$ in S/P-wave and P/S-wave, respectively. Assignment~2 is particularly unnatural in such models, as the $5/2^-$ state requires D-wave $\Sc \Db^*$ constituents and D-wave $\Jp p$ decay; it would be more natural to have $1/2^-$ or $3/2^-$ states, with S-wave constituents and S-wave decays.

An alternative interpretation of the opposite parities of the $\P\*$ doublet would be to regard one as the orbital excitation of the other. Unfortunately in such an approach the natural link to at least one of the thresholds is lost.

Pion-exchange is expected to be the dominant binding forced between flavoured hadrons. Karliner and Rosner~\cite{Karliner2015} predicted that since $\Lc\to \Lc\pi$ and $\Db\to\Db\pi$ couplings are forbidden by isospin and $J^P$ respectively, the lightest molecular state should appear at the $\Sc\Db^*$  threshold. Their predicted $I(J^P)=1/2(3/2^-)$ state is consistent with $\P^*$ in Assignment~3. The more general boson-exchange model of Yang~\etal.~\cite{Yang2011} finds the same state, as well as other S-wave states including one at $\Sc\Db$ threshold. He~\cite{He2015} considers both S- and P-wave interactions in the boson-exchange model, obtaining bound states in $\Sc^*\Db/\Sc\Db^*$ with quantum numbers consistent with Assignment~1. 

There are some models involving $\SD$ degrees of freedom which do not match the canonical identification of $\P/\P^*$ with $\Sc^*\Db/\Sc\Db^*$. In the local hidden-gauge approach, Xiao~\etal.~\cite{Xiao:2013yca} predicted a number of S-wave states across the $\SD$ sector, bound by around 50~MeV, and on this basis Roca~\etal.~\cite{arxiv-1507.04249} interpret $\P^*$ as a $\Sc\Db^*/\Sc^*\Db$ admixture, with $3/2^-$ quantum numbers as in Assignment~3. Chen~\etal.\cite{arxiv-1507.03704}, in a pion-exchange model, associate $\P/\P^*$ with higher-lying channels $\Sc\Db^*/\Sc^*\Db^*$, so the natural connection of their masses to threshold is lost, and as a pair of S-wave states their candidates cannot match the opposite parities of the experimental data. Using QCD sum rules Chen~\etal.~\cite{arxiv-1507.03717} associate the lighter state $\P$ with the heavier threshold $\Sc\Db^*$, and the heavier state $\P^*$ with the lighter threshold $\Sc^*\Db$.

The possibility of $\LD$ molecular states appears not to have been discussed in the literature. Naively $\Lc^*$ might not be expected in molecular states for the same reason that $\Lc$ is not: isospin forbids the coupling $\Lc\*\to \Lc\*\pi$. However this constraint disappears if one considers crossed channels, and a new possibility in this context is discussed in Sec.~\ref{molecule}.

Among the scenarios, the $\JpN$ baryocharmonium scenario, where $\P$ and $\P^*$ are $\Jp N(1440)$ and $\Jp N(1520)$ bound states~\cite{Kubarovsky:2015aaa}, is unique in allowing for S-wave couplings for both $\P\*$ states (Assignment~1). The model accounts naturally for the opposite parities of the states, and their mass gap. However it is not clear if the requisite binding, of the order of 150 MeV, can be realised in models~\cite{Kubarovsky:2015aaa}. A related possibility is that $\P^*$ is a baryocharmonium $\chip$ state. The problem with this is that S-wave coupling is not allowed for any of the $J^P$ assignments.

\section{Partner states}
\label{partners}

The models have different implications for the existence of partner states, yet to be discovered, with different isospin, charge and $J^P$: the observation or otherwise of these partners can help to discriminate among models.

Firstly, note that a feature which is presumably common to all models is the existence of neutral partners, decaying to $\Jp n$. This was noted in the context of the pentaquark interpretation by Lebed~\cite{Lebed:2015tna}. In the simplest picture, the neutral states are the $(I,I_3)=(1/2,-1/2)$ counterparts of the observed $(I,I_3)=(1/2,+1/2)$ states. Even if isospin is broken (discussed later) the neutral partners should exist.

Accurate determination of the masses of the neutral partners can help to reveal the underlying degrees of freedom. If $\P^*$ arises due to $\chip$ degrees of freedom then its neutral $\chi_{c1}n$ partner will presumably be heavier by around the $n$-$p$
mass difference,
\be
m_{n}-m_p=1.29\MeV.
\ee
On the other hand if $\P^*$ is due to $\LD$ degrees of freedom, its neutral partner has $\Db^{0}$ replaced with $D^{-}$, leading to a larger mass gap of several MeV,
\be
m_{D^-}-m_{\Db^0}=4.77\pm 0.08\MeV.
\ee
The situation is less conclusive in the $\SD$ scenario, because the charged $\P\*$ states and their neutral counterparts each have contributions from two charge combinations of meson-baryon pairs, as discussed in the next section.

Another striking difference among models is that unlike the $\chip$, $\LD$ and $\JpN$ scenarios,  the $\SD$ scenario, at least in principle, allows for $I=3/2$ partners. Observation of partner states with charge $+2$ and $-1$ with similar masses would be a signature of $\SD$ degrees of freedom. (Some relevant experimental channels are discussed in Sec.~\ref{decays}.) The models also have different implications for partners with other $J^P$, and a challenge for all models is to explain why the observed $J^P$ is unique.

In the model of Guo~\etal.~\cite{Guo2015} $\P^*$ is generated by $\chip$ rescattering and so will not have an $I=3/2$ partner at the $\chip$ threshold. (Possible rescattering effects at other thresholds, such as $\chi_{c1}\D$, should be investigated.) However (referring to Table~\ref{partials}) the P-wave $\chip$ interactions adopted in ref.~\cite{Guo2015} couple to $1/2^-$, $3/2^-$ and $5/2^-$; ideally the model should explain why the observed $J^P$ (either $1/2^-$ or $3/2^-$) is unique. Similarly, it would be interesting to know if their model generates $1/2^+$ or $3/2^+$ $\chip$ structures, given theoretical prejudice towards S-wave threshold effects.

In kinematic models where $\SD$ rescattering plays a role~\cite{Liu2015,Mikhasenko2015}, there is at least the possibility of $I=3/2$ partners for the $\P\*$ states; model calculations investigating kinematic peaks in the $\Jp\D$ mass spectrum would be welcome. In Assignment 1 (most natural in the $\SD$ scenario) the $3/2^-$ quantum numbers of $\P$ are unique in S-wave, so $1/2^-$ and $5/2^-$ partners are not expected. However the $5/2^+$ $\P^*$ could have $1/2^+$ or $3/2^+$ partners: it would be interesting if rescattering models can explain the apparent non-observation of such states in $\Jp p$. Similar remarks apply for other assignments.

Bound state models based on $\SD$ degrees of freedom differ considerably in their predictions for partners. Karliner and Rosner~\cite{Karliner2015} note that pion-exchange in the $\Sc\Db^*$ channel is equally attractive in $I(J^P)=1/2(3/2^-)$ and $3/2(1/2^-)$; if the former is identified with $\P^*$, a striking and simple consequence is the prediction of a degenerate $3/2(1/2^-)$ partner. (Their argument is based on a generalisation of the concept of relative binding number, or RBN, a numerical factor characterising the sign and magnitude of the interaction potential in different spin and isospin channels~\cite{Tornqvist:1991ks,Tornqvist:1993ng}. The computed values assume point-like pion-emission, but as shown in ref.~\cite{Burns:2014zfa}, the same values arise in other models for the pion vertex, including the $\an3P0$, flux tube, and microscopic models, so conclusions based on RBNs are rather general.)

Other models predict a richer spectroscopy. The boson-exchange model of Yang~\etal.~\cite{Yang2011} yields S-wave $\Sc\Db^*$ binding in all possible $I(J^P)$ channels, as well as $\Sc\Db$ binding in $3/2(1/2^-)$. In the local hidden-gauge approach Xiao~\etal.~\cite{Xiao:2013yca} find  deeply bound states in all possible $J^P$ channels formed from all possible $\SD$ combinations, but only in $I=1/2$.

The work of He~\cite{He2015} is unique in considering both S- and P-wave interactions, and restricts to $I=1/2$ channels. The $3/2^-$ $\S^*\Db$ state (Assignment~1) is found not to have a $3/2^+$ partner, but $1/2^+$ and $5/2^+$ possibilities are not discussed. The more striking result of ref.~\cite{He2015} is the P-wave $\Sc\Db^*$ binding in $5/2^+$, consistent with $\P^*$. As can be seen in Fig. 2 of that paper, some other channels are also bound with similar or smaller cut-offs, in particular the S-wave $3/2^-$ and P-wave $3/2^+$. The existence or otherwise of these partners is a test of model predictions.

If the $\P\*$ states are baryocharmonia built on $\JpN$ degrees of freedom,  the S-wave couplings (Assignment 1) lead to several possible $J^P$. Ideally the model should explain why the observed $3/2^-$ and $5/2^+$ states are preferred over the other possibilities $1/2^-$, $1/2^+$ and $3/2^+$. No $I=3/2$ partners are expected in the same mass region, though $\Jp\D$ partners might be possible. 

Note that the compact pentaquark interpretation (with quark, diquark, or triquark degrees of freedom) generally implies a proliferation of states with both $I=1/2$ and $I=3/2$, and several possible $J^P$: see, for example, Yuan~\etal.~\cite{Yuan:2012wz}. The number of states is particularly large because of the required orbital excitation, similar to the situation confronting the tetraquark model of $X(3872)$ in the case of the (now disproved) $2^{-+}$ assignment~\cite{Burns:2010qq}. The proliferation of states is less of a problem in the model of Lebed~\cite{Lebed:2015tna}, in which pentaquarks originating from $\Lb$ decay  have $I=1/2$ only, as  their $ud$ diquark is isoscalar since it is inherited from $\Lb$.

\section{Isospin violation}
\label{isospin}

An important aspect of the $\Sc^*\Db$ and $\Sc\Db^*$ channels has so far been overlooked in the literature. 
In any basis there are two ways to form a positive charge state from an isotriplet $\Sc^{(*)}$ and isodoublet $D^{(*)}$. In the charge basis, the two possibilities are distinguished by the third components of isospin,
\bea
\|\Sc^{(*)+}\Db^{(*)0}\>&=&|10,\thalf\thalf\>,\\
\|\Sc^{(*)++}\Db^{(*)-}\>&=&|11,\thalf-\thalf\>.
\eea
Note that the two possibilities have the same overall quark content, but differ in the arrangement of the quarks, namely $(udc)(u\c)$ and $(uuc)(d\c)$ respectively. Alternatively one can work in the basis of states of good total isospin, either 1/2 or 3/2,
\bea
\|(\Sc^{(*)}\Db^{(*)})_{\frac{1}{2},\frac{1}{2}}\>&=&|(1\ot\thalf)_{\frac{1}{2},\frac{1}{2}}\>,\\
\|(\Sc^{(*)}\Db^{(*)})_{\frac{3}{2},\frac{1}{2}}\>&=&|(1\ot\thalf)_{\frac{3}{2},\frac{1}{2}}\>.
\eea
The elements of the matrix translating between the bases are Clebsch-Gordan coefficients,
\be
\begin{array}{ccc}
										&\|\Sc^{(*)+}\Db^{(*)0}\>& \|\Sc^{(*)++}\Db^{(*)-}\>\\
\<(\Sc\* \Db\*)_{\frac{1}{2},\frac{1}{2}}\|	&-\sqrt{\frac{1}{3}}					&\sqrt{\frac{2}{3}}					\\
\<(\Sc\* \Db\*)_{\frac{3}{2},\frac{1}{2}}\|	&\sqrt{\frac{2}{3}}					&\sqrt{\frac{1}{3}}				\\
\label{cg}
\end{array}
\ee

Assuming the meson-baryon interactions respect isospin, if there were exact degeneracy within each ($\Sc^{(*)+}$, $\Sc^{(*)++}$) and ($\Db^{(*)0}$, $\Db^{(*)-}$) pair, the physical states would be $I=1/2$ and $I=3/2$ eigenstates as determined by the above mixing matrix. But this limit is not realised in nature: the thresholds are, for $\Sc^* \Db$,
\bea
M(\Sc^{*+})+M(\Db^0)&=&4382.3\pm 2.4\MeV,\\
M(\Sc^{*++})+M(D^-)&=&4387.5\pm 0.7\MeV,
\eea
and for $\Sc\Db^*$,
\bea
M(\Sc^+)+M(\Db^{*0})&=&4459.9\pm 0.5\MeV,\\
M(\Sc^{++})+M(D^{*-})&=&4464.24\pm 0.23\MeV.
\eea
The mass gap of around 5 MeV for each pair of thresholds, while small on typical hadronic scales, is significant compared to the binding energies of $\P$ and $\P^*$. The result is that in each case the lower-lying $\Sc^{(*)+}\Db^{(*)0}$ component of the wavefunction will be enhanced compared to the $\Sc^{(*)++}\Db^{(*)-}$ component. The physical states are not states of good isospin, then, but admixtures of $I=1/2$ and $I=3/2$. In the extreme case that the $\Sc^{(*)++}\Db^{(*)-}$ component is negligible, the probability that the physical state is in $I=3/2$ is twice that of $I=1/2$.

The situation is analogous to the case of $X(3872)$, where the observed isospin violation is understood in terms of the mass gap separating the $D^{*0}\Db^0$ and $D^{*+}\Db^-$ channels~\cite{Close:2003sg,Tornqvist:2004qy,Suzuki:2005ha,Gamermann:2009fv,Burns:2011uy}. For $X(3872)$, isospin breaking is scale-dependent, and is larger at large distances~\cite{Voloshin:2006wf,Gamermann:2009uq}. The same is expected here, meaning isospin effects differ in various production and decay processes. In any case, isospin violation will be present at some level, and will have observable consequences.

Note that this isospin violation is a feature of the $\SD$ scenario generally, appearing in both kinematic and bound state models. In kinematic models the enhancement of $\Sc^{(*)+}\Db^{(*)0}$ is associated with its energy denominator in the loop integral. In bound state models (at least in S-wave) it is evident in the universal wavefunction applicable to loosely-bound states: see for example refs~\cite{Voloshin:2006wf,Gamermann:2009uq}. 

Due to their $\Jp p$ decays, it would be natural to assign the $\P\*$ states to $I=1/2$ doublets. But if $\SD$ degrees of freedom are playing a role the required isospin violation implies that this canonical interpretation no longer applies. Two distinct and interesting possibilities arise. The closest match to the canonical interpretation is to place the states in putative $I=1/2$ doublets, meaning that, were it not for the mass splittings of their constituents, they would have $I=1/2$. But there is a novel and equally plausible alternative: they could be putative $I=3/2$ states, and their observed $\Jp p$ decays are actually a manifestation of the required isospin breaking.

A striking confirmation of the latter interpretation would be the observation of $(I,I_3)=(3/2,\pm 3/2)$ partners to the $\P\*$ states, with charge $+2$ or $-1$. However it is not automatic that such states will be bound. If the mass of the observed $\P\*$ state(s) includes a downward contribution due to mixing of $I=3/2$ and $I=1/2$, the $(3/2,\pm 3/2)$ partners will be somewhat heavier and not necessarily bound.

Regardless of whether the $\P\*$ states arise from putative $I=1/2$ or $I=3/2$ doublets, if $\SD$ interactions play a role they will have mixed isospin and their production and decays will reflect this. By contrast, in the $\chip$, $\LD$ and $\JpN$ scenarios the $\P\*$ states have $I=1/2$.

\section{Decay patterns} 
\label{decays}

Experimental observation of the $\P\*$ states in various decay modes can discriminate among the possible meson-baryon degrees of freedom, as the expected decay patterns differ for the various scenarios.

On general grounds, many decays other than the observed $\Jp p$ can be expected. The kinematically accessible two-body modes are the open-charm pairs $\Lc\Db$, $\Lc\Db^*$, $\Sc\Db$ and (for $\P^*$) $\Sc^*\Db$, and closed-charm pairs $\eta_c p$, $\chi_{c0} p$ and (if isospin is broken) $\Jp\D$ and $\eta_c \D$. In addition, there are several three-body channels of interest: $\Jp N\pi$, $\Lc\Db\pi$ and (for $\P^*$) $\Lc\Db^*\pi$ and $\Sc^+\Db^0\pi^0$. (Note that for $\Sc \Db\pi$ only the specified charge channel is kinematically accessible, and only due to the finite width of $\P^*$.)

The aim of this section is to distinguish which channels are and are not available in each scenario, based only on the assumed spin and flavour degrees of freedom. Whether or not the decays allowed by these arguments translate into prominent decays in specific models requires more detailed calculations, beyond the scope of this paper. For example, if the $\P\*$ states are purely kinematic effects then it is possible that they will not be seen in any channels other than the observed $\Jp p$: only detailed model calculations can establish this. Moreover, decays allowed by the arguments below may turn out to be small due to partial wave suppression; once the experimental $J^P$ assignments are determined definitively, the summary of partial waves in Table~\ref{partials} can be used as a guide.

However, there is an indirect argument which suggests that significant decays other than $\Jp p$ may be expected. Wang~\etal.~\cite{Wang:2015jsa} argue that if the $\P\*$ states are resonances and not kinematic effects, they should be seen in $ \gamma p\to \Jp p$, and that in this case, existing experimental data require that the $\P\*\to\Jp p$ branching fractions are small.

As usual, the analysis in this section treats the competing scenarios as distinct. Even if this is too simple a picture, and the $\P\*$ states involve some interplay among different meson-baryon degrees of freedom, the conclusions can still be useful. If the $\P\*$ states are eventually observed in several channels which are not all allowed within a given scenario, it could indicate the presence of mixed degrees of freedom; an example is given in Sec.~\ref{molecule}.

\begin{table}
\caption{Predictions for allowed (\t) and suppressed (\x) decays for the different scenarios. The absence of an entry implies that a given channel is not kinematically accessible. The predictions enclosed in brackets are less reliable and can be badly violated if pion-exchange dominates: see the text.}
\label{dec}       
\begin{tabular}{p{1cm}p{0.7cm}p{0.7cm}p{0.7cm}p{0.7cm}p{0.1cm}p{0.7cm}p{0.7cm}}
\hline\noalign{\smallskip}
			&\multicolumn{4}{c}{$\P^*$}								&&\multicolumn{2}{c}{$\P$}\\
\noalign{\smallskip}
\cline{2-5}\cline{7-8}
\noalign{\smallskip}
			&$\chip$	&$\SDh$		&$\LD$		&$\JpN$				&&$\SDl$		&$\JpN$\\
\noalign{\smallskip}\hline\noalign{\smallskip}
$\Jp N$		&\t			&\t			&\t			&\t					&&\t			&\t				\\
$\eta_c N$	&\x			&\x			&\t			&\x					&&\x			&\x				\\
\noalign{\smallskip}\hline\noalign{\smallskip}
$\Jp \D$	&\x			&\t			&\x			&\x					&&\t			&\x				\\
$\eta_c \D$	&\x			&\t			&\x			&\x					&&\t			&\x				\\
\noalign{\smallskip}\hline\noalign{\smallskip}
$\Lc\Db$	&\t			&[\x]		&[\t]		&\x					&&[\x]			&\x				\\
$\Lc\Db^*$	&\t			&\t			&[\t]		&\t					&&\t			&\t				\\
$\Sc\Db$	&\t			&[\x]		&\t			&\x					&&[\x]			&\x				\\
$\Sc^*\Db$	&\t			&\t			&[\x]		&\t					&&			&				\\
\noalign{\smallskip}\hline\noalign{\smallskip}
$\Jp N\pi$	&\x			&\t			&\x			&\t					&&\t			&\t				\\
$\Lc \Db\pi$&\x			&\x			&\x			&\x					&&\t			&\x				\\
$\Lc \Db^*\pi$&\x		&\t			&\x			&\x					&&			&				\\
$\Sc^+\Db^0\pi^0$&\x		&\t			&\t			&\x					&&			&				\\
\noalign{\smallskip}\hline
\end{tabular}
\end{table}

The patterns of strong decays expected for the different scenarios are summarised in Table~\ref{dec}, and explained below. For convenience, charge labels are dropped (except for $\Sc^+\Db^0\pi^0$) and the label $N$ will be used to stand for the $(n,p)$ isodoublet: the conclusions below apply both to the charged $\P\*$ states and their neutral partners.

Isospin leads to a simple selection rule: the $\Jp\D$ and $\eta_c\D$ modes have $I=3/2$, so are only possible in the $\SD$ scenario, where isospin is broken. Otherwise all channels are allowed by isospin in all of the scenarios.

Another strong constraint comes from heavy-quark spin conservation. This can only be applied to transitions in which both the initial and final heavy-quark spins are fixed, namely for transitions between closed charm states $(uud)(\cc)\to (uud)(\cc)$.

If either the initial or final state has open charm, its heavy-quark spin is not fixed. In such cases it will be assumed that total (heavy plus light) quark spin is conserved. This leads to selection rules for certain final states which have unique quark spins $S=1/2$ or $S=3/2$. In this picture transitions between open- and closed-charm states $(udc)(\uc)\leftrightarrow (udu)(c\c)$ are the result of $u/c$ quark interchange, and those between open- and open-charm states $(udc)(\uc)\leftrightarrow(udc)(\uc)$ and $(udc)(\uc)\leftrightarrow(uuc)(\dc)$ are the result of $u/u$ or $u/d$ interchange. The assumption is that the quark-level interactions which cause transition from initial to final state conserve spin. This is true of models, such as refs \cite{Swanson:2003tb,Vijande:2007fc}, in which hadron-hadron interactions are derived from the dominant quark-quark interactions that determine the hadron spectrum, namely long-range confinement and short-range Coulomb attraction (which are spin-independent) and hyperfine interactions (which are spin-dependent, but conserve total spin).
 
Instead of the interchange of quarks, these decays can also be described in terms of the exchange of mesons, with $\uc$ flavour for
open- to closed-charm transitions, or $\uu$/$\ud$ flavour for open- to open-charm transitions. While topologically equivalent in terms of quark line diagrams, the quark interchange and meson exchange processes use different interaction potentials. The meson-level description will presumably be more important when the exchanged meson can be a pion, namely for the open- to open-charm transitions. In such cases the requirement of total spin conservation should not be taken too seriously, since in the deuteron, for example, tensor pion-exchange interactions (which do not conserve spin) are important \cite{Ericson:1993wy}.

Predictions for three-body decays are of course more difficult. These are assumed to be substantial only if they are accessible through a single vertex, or an intermediary with large decay width. Finally, note that the $\chi_{c0}p$ mode is not included in the table or discussion below, since on the basis of arguments presented here it does not discriminate among the models.

With reference to Table~\ref{dec}, predictions for the different scenarios will now be discussed in turn. In the $\chip$ scenario the $c\c$ constituents of $\P^*$ are coupled to spin $S_{c\c}=1$. Assuming heavy-quark spin conservation, the $\eta_c N$ mode is therefore not available, since it has $S_{c\c}=0$. Both $\Jp\D$ and $\eta_c\D$ are forbidden by isospin, and the latter is also forbidden by heavy-quark spin.

The remaining two-body modes are $\Lc\Db\*$ and $\Sc\*\Db$, via transitions $(qqq)(c\c)\to (qqc)(q\c)$. A simple recoupling calculation shows that each of  $\Lc\Db\*$ and $\Sc\*\Db$ is an admixture of $S_{c\c}=0$ and $S_{c\c}=1$, so all of these channels are allowed by heavy-quark spin conservation. There are no additional constraints from the conservation of total quark spin, since $\chip$ can have $S=1/2$ or $S=3/2$. Three-body modes $\Jp N\pi$, $\Lc \Db\*\pi$ and $\Sc^+\Db^0\pi^0$ are higher order and will presumably be suppressed. Guo~\etal.~\cite{Guo2015} note that observation of $\P^*$ in $\chip$ itself would identify it as a genuine composite, as opposed to a kinematic effect.

In the $\SD$ scenario,  modes involving $\D$ are accessible due to isospin mixing. As $\SDh$ and $\SDl$ are admixtures of $S_{c\c}=0$ and $S_{c\c}=1$, the $(qqc)(q\c)\to(qqq)(c\c)$ transition responsible for the observed $\Jp N$ can also, according to heavy quark spin, access $\eta_c N$. (The situation is similar to the molecular model for the $Z_b$ states~\cite{Bondar:2011ev}.)  However there are additional constraints from the conservation of total quark spin. The $\SDl$ combination has $S=3/2$, and while $\SDh$ can in general have $S=1/2$ or $S=3/2$, for Assignments 1 and 3 (the most natural assignments for the $\SD$ scenario) only $S=3/2$ is possible (this is required to form $5/2^+$ in P-wave, or $3/2^-$ in S-wave: see Table~\ref{partials}). Assuming only $S=3/2$ transitions are allowed, the $\eta_c N$ mode ($S=1/2$) is forbidden. By contrast, $\Jp N$, $\Jp\D$ and $\eta_c\D$ are allowed. The $\Jp N\pi$ mode (fed by the broad $\D$ in $\Jp \D$) is also expected. (There may be additional contributions from intermediate $\JpN$ states.) The $\eta_c N\pi$ mode (not shown in the table) is also likely for the same reason.

(In the next section, a simple model will be used to predict the branching fractions for the $\Jp\D$ and $\eta_c\D$ modes. It will also be shown that, even if the broad $\D$ cannot be resolved explicitly in experimental analysis, it can be inferred indirectly from the charge combinations in the $\Jp N\pi$ mode.)

For the open-charm to open-charm transitions the requirement that $S=3/2$ is conserved implies $\Lc\Db$ and $\Sc\Db$ are forbidden, while $\Lc\Db^*$ and (for $\P^*$) $\Sc^*\Db$ are allowed. However these selection rules do not apply if spin-violating pion-exchange transitions are allowed. The known decays $\Sc\*\to \Lc\pi$ imply that, as $\SDl/\SDh$ states, $\P/\P^*$ will decay to $\Lc \Db\pi/\Lc \Db^*\pi$, respectively. Note that $\P^*$ is not expected in $\Lc\Db\pi$ despite its being kinematically allowed. Finally for $\P^*$ the $\Sc^+\Db^0\pi^0$ mode is accessible by the decay of the $\Db^*$ component in $\SDh$.

In the $\LD$ scenario $\P^*$ is an $I=1/2$ state, and so cannot decay into $\Jp\D$ or $\eta_c\D$. Since $\Jp\D$ is forbidden the multi-body $\Jp N\pi$ mode is also expected to be small, although there may be contributions (difficult to estimate) from intermediate $\JpN$ states. The three-body mode $\Sc^+\Db^0\pi^0$ is expected due to the strong S-wave decay $\Lc^*\to\Sc^+\pi^0$, whereas $\Lc\Db\pi$ is forbidden because isospin forbids $\Lc^*\to\Lc\pi$.

For the $\LD$ scenario, selection rules based on total quark spin require some remarks on the spin wavefunction of $\Lc^*$. As the $ud$ flavour wavefunction is antisymmetric (isoscalar), its spin and spatial symmetries are overall antisymmetric. There are three ways to form P-wave baryon with $1/2^-$ quantum numbers: exciting the  $\lambda$ coordinate (between the $ud$ centre of mass and $c$) gives $ud$ spin 0 and total quark spin 1/2; exciting the $\rho$ coordinate (between $u$ and $d$) gives $ud$ spin 1 and total quark spin 1/2 or 3/2. In principle the true wavefunction is a linear combination of all three, however in most models the first component dominates. In the quark-diquark model this is assumed {\it a~priori} \cite{Ebert:2007nw}. In potential models the $\lambda$ excitation is lowest in energy, and so dominates \cite{Copley:1979wj}. In the relativistic quark model the dominance of $S=1/2$  is demonstrated explicitly in ref. \cite{Migura:2006ep}. 

Given that the quark spin of $\Lc^*$ is (dominantly) 1/2, the $\LD$ combination also has total quark spin $S=1/2$. This means that the $\eta_c N$ channel is accessible, unlike in other scenarios. Spin conservation also allows $\Lc\Db\*$ and $\Sc\Db$, but (assuming $S=1/2$) not $\Sc^*\Db$. However if pion exchange is responsible for any open- to open-flavour decays, the pattern reverses: the $\Lc\Db\*$ and $\Sc\Db$ transitions are forbidden since $\Lc^*\to\Lc\pi$ is forbidden by isospin, while $\Sc^*\Db$ is allowed by spin-violating pion transitions. Quark spin conservation also has implications for the observed $\Jp N$ mode. For Assignment~1 this is P-wave in general, but F-wave if $S=1/2$ is required, and in Assignment~3 this is S-wave in general, but D-wave if $S=1/2$ is required; these constraints are not a feature of the $\SD$ scenario.

In the $\JpN$ scenario, heavy-quark spin conservation forbids $\eta_c N$, and isospin forbids $\Jp\D$ and $\eta_c\D$. The internal decay $N^*\to N\pi$ will lead to significant $\Jp N\pi$~\cite{Kubarovsky:2015aaa}, but other multi-body modes are higher order and will presumably be small. Total quark spin conservation gives additional constraints. Recall that the $\JpN$ scenario fits Assignment~1 with S-wave couplings. In this case the $3/2^-$ quantum numbers of $\P$ require $\Jp N(1440)$ to have total quark spin $S=3/2$. For $\P^*£$, note that $N(1520)$ is a P-wave baryon whose quark spin wavefunction is dominantly $S=1/2$, with a small $S=3/2$ admixture \cite{Agashe:2014kda}; ignoring the latter, to form $5/2^+$ quantum numbers the total  quark spin of $\Jp N(1520)$ is $S=3/2$. Hence spin conservation requires $S=3/2$ final states for both $\P$ and $\P^*$ in the $\JpN$ scenarios: for open charm modes, $\Lc\Db^*$ and (for $\P^*$) $\SDl$ will dominate.

Finally, note that some radiative decays can also be expected. Given the coupling of $\Jp$ to the photon, the decays $\Jp N$, $\Jp\D$ and $\Jp N\pi$ allowed in Table~\ref{dec} imply associated decays $\gamma N$, $\gamma\D$ and $\gamma N\pi$. Other channels are possible via the radiative decay of one of the meson-baryon constituents, such as the $\SDh$ state decaying into $\Sc\Db\gamma$.

\section{Relations among decays}
\label{relations}
In this section some relations among decays in different channels are discussed. A simple model is used to predict the missing $\Jp\D$, $\eta_c\D$ and $\eta_c N$ modes, and it is shown that for other modes, the relative strengths of different charge combinations can indicate the isospin of $\P\*$ states.

Whereas absolute predictions for decay widths are highly model-dependent, relations among decays are more general. In order to make predictions for missing modes, we are interested in relations which include $\Jp N$, the only mode so far observed. The channels which can be related to this involve hadrons with essentially the same spatial wavefunctions but different quark spin and flavour wavefunctions, namely $\Jp\D$, $\eta_c\D$ and $\eta_c N$. These decays (see Table~\ref{dec}) are expected in the $\SD$ and $\LD$ scenarios.

So we are interested in transitions of the form $(qqc)(q\c)\to (qqq)(c\c)$. As noted previously, such transitions can be realised either in terms of the interchange of a $c$ and a $q$ quark, or the exchange of mesons with $q\c$ flavour. In this paper the former approach will be adopted; although the two pictures differ somewhat, for present purposes their essential conclusions are the same, as will be shown later. 

In quark-exchange models, the transition between two hadronic channels is typically triggered by interactions between the quark constituents of the different hadrons. It will be assumed here that the interaction potential is spin-independent. Although models often include spin-dependent hyperfine interactions, it is reasonable to assume that these are smaller than the spin-independent (confinement and Coulomb) terms. For the S-wave coupling between open- and closed-charm mesons this was found explicitly in ref. \cite{Swanson:2003tb}. As hyperfine interactions are short range, the assumption is even safer when initial or final state involves non-zero partial waves, as is the case for many possible $\P\*$ transitions.

The decay amplitude factorises into separate amplitudes for the quark spin, flavour and spatial degrees of freedom. Assuming a spin-independent interaction, the combined spin and flavour amplitude is just overlap of the spin-flavour wavefunctions of the initial and final states after exchanging a $c$ and a $q$ quark; these are straightforward, and are discussed below. The spatial part is an integral over the radial and orbital wavefunctions, modulated by the interaction potential, and depends on the decay momentum $p$ and the partial wave $l$. Since the final states are in each case coupled to the same quark spin, their decays are accessible in the same partial wave $l$. In the limit of identical spatial wavefunctions for $\Jp$ and $\eta_c$, and for $N$ and $\D$, the functional form of the spatial amplitude is common to the different channels, but its numerical value differs because of the different decay momenta. Explicit calculation of the spatial amplitude is model-dependent. An alternative would be to assume the near-threshold behaviour $p^l$, but until the experimental $J^P$ assignments are established definitively the partial waves $l$ are not known. For simplicity, in this paper the spatial part will be treated as a common factor for the different channels. A similar approach has been applied to meson strong decays by quark-pair creation~\cite{Burns:2006wz,Burns:2007hk,Burns:2013xoa,Burns:2014zfa}.

Assuming a common spatial part, the decay amplitudes for the various channels are proportional to spin-flavour wavefunction overlaps. These are easily obtained, so details of the calculation will not be given here. (As noted later, the computed overlaps are consistent with related calculations elsewhere in the literature.) 

For the $\SDl$ state $\P$, the overlaps ($S=3/2$) are
\bea
\< \Jp N \| (\SDl)_{\frac{1}{2}}\>&=&-\frac{1}{\sqrt 6},\\
\< \Jp \D \| (\SDl)_{\frac{3}{2}}\>&=&\frac{1}{2}\sqrt{\frac{5}{3}},\\
\< \eta_c \D \| (\SDl)_{\frac{3}{2}}\>&=&\frac{1}{2},
\eea
where the subscripts specify the isospin. Notice that the overlaps in the $I=3/2$ channel are large. As $\P$ is a state of mixed isospin, consider a fixed mixing angle
\be
\| \P\>=\cos\phi \| (\SDl)_{\frac{1}{2}}\>
+\sin\phi \| (\SDl)_{\frac{3}{2}}\>,
\label{pmixing}
\ee
so that, before phase space factors, the branching fractions are related
\be
\Jp N~:~\Jp \D~:~ \eta_c\D = 
2\cos^2\phi ~:~5\sin^2\phi ~:~ 3 \sin^2\phi.
\ee
The $\Jp\D$ and $\eta_c\D$ decays are suppressed due to phase space factors, but are enhanced by the larger wavefunction overlaps. If the mixing angle is small then these modes may not be observable, but as discussed in Sec.~\ref{isospin}, there are good arguments in favour of a large mixing. Firstly, in the limit that the lowest lying charge state entirely dominates the wavefunction, $\sin^2\!\phi = 2 \cos^2\!\phi$. Secondly, it is entirely possible that $\P$ is a putative $I=3/2$ state whose $\Jp p$ mode is a manifestation of isospin breaking: in this case $\sin^2\!\phi$ will be large compared to $\cos^2\!\phi$.

Admittedly the choice of a fixed mixing angle is too simplistic: the mixing is momentum-dependent (see Sec.~\ref{isospin}) and in a more detailed calculation this should be included in the spatial amplitude. But the general conclusion remains: if $\P$ is a $\SDl$ state then its $\Jp\D$ and $\eta_c\D$ decays ought to be significant.

For the $\SDh$ state $\P^*$, in general two spin states are possible, but the most natural interpretations (as remarked in the previous section) have $S=3/2$. In this case the overlaps are
\bea
\< \Jp N \| (\SDh)_{\frac{1}{2}}\>&=&\frac{1}{3\sqrt 2},\\
\< \Jp \D \| (\SDh)_{\frac{3}{2}}\>&=&{\frac{\sqrt 5}{3}},\\
\< \eta_c \D \| (\SDh)_{\frac{3}{2}}\>&=&-\frac{1}{\sqrt 3},
\eea
so that with mixing
\be
\| \P^*\>=\cos\phi \| (\SDh)_{\frac{1}{2}}\>
+\sin\phi \| (\SDh)_{\frac{3}{2}}\>,
\label{P*mixing}
\ee
the  branching fractions are related
\be
\Jp N~:~\Jp \D~:~ \eta_c\D = 
\cos^2\phi ~:~ 10\sin^2\phi ~:~ 6\sin^2\phi.
\label{bf1}
\ee
Even more than in the case of $\P$, the $I=3/2$ decay modes of $\P^*$ are enhanced by large numerical factors. The same arguments for the possible large mixing angle apply, and these modes should therefore be substantial.

The overlaps above can be applied immediately to the $(I,I_3)=(3/2,\pm 3/2)$ partners of the $\P\*$ states, if they exist. The $\P^{++,-}$ and $\P^{*++,-}$ states will decay into $\Jp \D^{++,-}$ and $\eta_c \D^{++,-}$, with the same relative branching fractions,
\be
\Jp\D~:~\eta_c\D = 5~:~ 3.
\label{bf2}
\ee
Note that the $\P^{(*)++,-}$ states are likely to be narrower than the $\P^{(*)+}$ states, since they have fewer available decay modes -- in particular the $\Jp p$ mode is not available -- and their allowed modes with $\D^{++,-}$ have less phase space.

In the $\LD$ scenario, assuming $S=1/2$ as described in the previous section, the overlaps are
\bea
\< \Jp N \| (\LD)_{\frac{1}{2}}\>&=&\frac{1}{2}\sqrt{\frac{3}{2}},\\
\< \eta_c N \| (\LD)_{\frac{1}{2}}\>&=&{\frac{1}{2\sqrt 2}},
\eea
so that, before correcting for phase space and the spatial amplitude, the branching fractions are
\be
\Jp N~:~ \eta_c N=
3 ~:~1.
\label{bf3}
\ee

It is worth noting how these results compare to others in the literature. Garc\'ia-Recio~\etal.~\cite{Garcia-Recio:2013gaa} computed the masses of baryons with hidden charm in a coupled-channel model with spin-flavour symmetry. All of the overlaps quoted above are consistent (in relative magnitude) with their quark interchange matrix elements. The coupled-channel model of Xiao~\etal.~\cite{Xiao:2013yca} differs in treating transitions not as the result of the interchange of quarks, but the exchange of $D^*/\Db^*$ mesons. Their matrix elements are,  for a given isospin channel, consistent in relative magnitude with all of the above, but the relative weight of the $I=3/2$ and $I=1/2$ channels differs. Their $I=3/2$ amplitudes are enhanced relative to the above by a factor of $3/2$ in amplitude (9/4 in branching fraction), so in their model the $\Jp\D$ and $\eta_c\D$ modes would be even more prominent.

Once the experimental $J^P$ quantum numbers of the $\P\*$ states are determined definitively, the above relations can be improved using the known decay partial wave $l$, by scaling the spatial matrix elements according to the threshold $p^l$ behaviour (and including phase space factors). These relations can be combined with the recently-obtained product branching fractions $\mathcal B(\Lb\to\P^{(*)+}K^-)\times \mathcal B(\P^{(*)+}\to\Jp p)$  \cite{Aaij:2015fea} to predict the corresponding product branching fractions for $\Jp\D$, $\eta_c\D$ and $\eta_c N$.

For modes in which several charge combinations are possible (namely $\Sc\*\Db$, $\Jp N\pi$, and $\Lc\Db\*\pi$) the relative branching fractions are controlled by isospin factors. This can be used to distinguish among models. For the discussion below, refer to Table~\ref{dec}.

The $\Sc\Db$ decay for $\P^*$ is possible in both the $\chip$ and $\LD$ scenarios. As pure $I=1/2$ states we expect, from equation~\rf{cg}, the relative branching fractions 
\be
\Sc^+\Db^0~:~\Sc^{++}D^-=1~:~2.
\label{bf4}
\ee

The $\Sc^*\Db$ decay could be more revealing. In the $\chip$ and $\JpN$ scenarios we expect, as above,
\be
\Sc^{*+}\Db^0~:~\Sc^{*++}D^-=1~:~2.
\label{bf5}
\ee
On the other hand in the $\SDh$ scenario some isospin violation is expected, and the branching fractions will deviate from the above. With the mixing as defined in equation~\rf{P*mixing}, the contributions of the different charge combinations are, from equation \rf{cg},
\be
\|\P^*\>
=\sin(\phi-35.3^\circ)
\|\Sc^{*+}\Db^0\>
+\cos(\phi-35.3^\circ)
\|\Sc^{*++}D^-\>
\ee
using $\sin35.3^\circ=\sqrt{1/3}$ and $\cos35.3^\circ=\sqrt{2/3}$. This implies the relative branching fractions
\be
\Sc^{*+}\Db^0~:~\Sc^{*++}D^-=\sin^2(\phi-35.3^\circ)~:~\cos^2(\phi-35.3^\circ),
\label{bf6}
\ee
which reduces to the $1:2$ ratio expected for an $I=1/2$ state in the absence of mixing.
 
Turning now to the three-body modes, note that $\Jp N\pi$ is expected for $\P$ and $\P^*$ in both the $\JpN$ and $\SD$ scenarios. In the former case this is due to the decay of the constituent $N^*$, so the relative branching fractions follow from $I=1/2$ Clebsch-Gordan coefficients:
\be
\Jp n \pi^+ ~:~ \Jp p\pi^0=2~:~ 1.
\label{bf7}
\ee
On the other hand in the $\SD$ scenario this decay will presumably be dominated by $\Jp \D$ with $\D\to N\pi$. Ignoring contributions from decays through $N^*$ states (which are difficult to estimate, but likely to be smaller) implies the opposite pattern to the above
\be
\Jp n \pi^+ ~:~ \Jp p\pi^0=1~:~ 2.
\label{bf8}
\ee
This is a clean signature for the presence of $I=3/2$ components in the $\P\*$ states, which is particularly useful given the likely experimental difficulty of identifying the broad $\D$ explicitly in $\Jp\D$ decays.

Finally the $\Lc\Db\*\pi$ modes are expected only in the $\SD$ scenario. Some isospin violation is expected here, and with mixing defined as in eqns~\rf{pmixing} and~\rf{P*mixing} the branching fractions are
\begin{multline}
\Lc^+\Db^{(*)0}\pi^0~:~\Lc^+D^{(*)-}\pi^+=\\\sin^2(\phi-35.3^\circ)~:~\cos^2(\phi-35.3^\circ),
\label{bf9}
\end{multline}
reducing to the $1:2$ ratio expected for an $I=1/2$ state in the absence of mixing. (Mass differences among the neutral and charged states will modify these predictions somewhat, particularly for the $\P^*\to \Lc \Db^*\pi$ decays, which have little phase space.)

\section{Production}
\label{production}

In this section some new production modes for $\P\*$ states and their possible partners are discussed.

Starting from $\Lb$, the Cabibbo-favoured $b\to s \cc$ transition combined with the creation of a light quark $\nn$ pair yields a final state with quark content $uds \cc\nn$. From the isospin zero combination $\nn=(\uu+\dd)/\sqrt{2}$, the $\uu$ component gives the observed $\Lb\to \Jp p K^-$ decay and the $\P^{(*)+}$ states, while the $\dd$ component leads to $\Lb\to \Jp n \bar K^0$ and should give a comparable yield of the neutral partner states $\P^{(*)0}$~\cite{Lebed:2015tna}.

As discussed in Sec.~\ref{decays}, the $\P\*$ states may decay in many modes other than $\Jp N$. This suggests experimental study of $\Lb\to \P^{(*)+}K^-$ and $\Lb\to \P^{(*)0}\bar K^0$, with $\P\*$ observed in any of the two- or three-body modes listed in Table~\ref{dec}. 

The discovery of the states in $\Lb\to \Jp p K^{-}$ suggests experimental study of $\Lb\to \Jp p K^{*-}$, where comparable yields could be expected. Model predictions for the relative yields would be a useful discriminator. For example, kinematically the $ \Jp p K^{*-}$ and $ \Jp p K^{-}$ modes differ, so in models based on kinematic singularities it is not obvious whether or not the $\P\*$ states are expected in both.

The Cabibbo-suppressed $b\to d\cc$ transition leads to a final state with flavour $udd \cc\nn$. The $\P^{(*)+}$ states could therefore be observed in $\Lb\to \Jp p \pi^-$~\cite{Hsiao2015}, and the $\P^{(*)0}$ states in $\Lb\to \Jp n \pi^0$. In this context there is an interesting experimental fact which has not been noted in the literature. Prior to the discovery of $\P\*$ states, the LHCb Collaboration reported on the observation of the $\Lb\to \Jp p \pi^-$ decay, and noted that there is no sign of exotic structures in the $\Jp p$ mass spectrum~\cite{Aaij:2014zoa}. This was not discussed in the  $\P\*$ discovery paper \cite{Aaij:2015tga}. 

Whether or not the absence of a signal is statistically significant remains to be seen: the  $\Lb\to \Jp p \pi^-$ branching fraction is around a factor of ten smaller than $\Lb\to \Jp p K^-$, for example. Intriguingly there is a peak in the $\Jp p$ invariant mass plot in ref.~\cite{Aaij:2014zoa} around 4.4~GeV (see their Fig.~2). Experimental analysis of the $\Lb\to \Jp p \pi^-$ mode would be very useful: observation of the $\P\*$ states, or an upper limit on their production, would help to discriminate among their possible interpretations. For this purpose, theoretical estimates of the relative yields of $\P\*$ states in the different production modes are required. Already there is one such prediction in the literature: in the model of ref.~\cite{Hsiao2015},
\be
\frac{\mathcal{B}(\Lb\to \P^{(*)+} \pi^-)}{\mathcal{B}(\Lb\to \P^{(*)+} K^-)}=0.8\pm 0.1.
\ee

The Cabibbo-suppressed transition is also interesting as an indicator of isospin. Whereas the Cabibbo-favoured transition can only produce states with flavour $uud\cc$ and $udd\cc$, the Cabibbo-suppressed transition can additionally produce the $ddd\cc$ state. In some models only the first two states (an isodoublet) are expected. In the $\SD$ scenario, however, the $\P\*$ states could be putative $I=3/2$ states, or could have $I=3/2$ partners. The Cabibbo-suppressed transition gives a unique possibility to produce the negatively charged $(I,I_3)=(3/2,-3/2)$ state $ddd\cc$. 

As the $ud$ pair in $\Lb$ is isosinglet, the $udd \cc\nn$ combination has $I=1/2$. If the $\P\*$ states (or their partners) are pure $I=1/2$, only two charge modes are expected, and their relative production branching fractions are determined by $1/2\to 1/2\times 1$ Clebsch-Gordan coefficients,
\be
\P^{(*)+}\pi^-~:~\P^{(*)0}\pi^0 = 2~:~1.
\ee
Discrepancies from the above ratio would indicate isospin violation, consistent with expectations for the $\SD$ scenario. An $I=3/2$ multiplet is characterised by three charge modes, weighted by $1/2\to 3/2\times 1$ Clebsch-Gordan coefficients,
\be
\P^{(*)+}\pi^-~:~\P^{(*)0}\pi^0~:~\P^{(*)-}\pi^+ = 1~:~2~:~3.
\ee
Note that the negatively charged state, which is a clean signature of an $I=3/2$ multiplet, is produced most copiously.

In addition to $\Lb$ decays, there are some other possible sources of $\P\*$ and related states. As noted by several authors~\cite{Karliner2015a,Kubarovsky:2015aaa,Wang:2015jsa},  if they  are genuine resonances and not kinematic effects, the $\P\*$ states should be observable in $\gamma p\to \Jp p$. Likewise they could be found in $\gamma p$ photoproduction in any of the final states described in Table~\ref{dec}.

In $\Upsilon(1S)$ decays and $\epem$ the $\P\*$ states could be observed in the $\Jp p \bar p$ mode~\cite{Guo2015,Wang:2015jsa}. Similarly we could expect the same processes to produce $\P^{(*)+}\bar p$, with $\P^{(*)+}$ observed in any of the modes in Table~\ref{dec}. If the $\P\*$ states have mixed isospin, or have $I=3/2$ partners, the same processes would produce $\P\*\bar \D$.

Prior to the discovery of the $\P\*$ states, Wu~\etal.~\cite{Wu2010} noted that hidden charm baryons could be studied at the PANDA experiment, in $p\bar p\to \Jp p\bar p$ and $p\bar p\to \eta_c p\bar p$. With a 15 GeV $\bar p$ beam the states are just energetically-allowed; there is, however, insufficient energy to produce the $I=3/2$ partners recoiling against $\bar \D$. 

Note that the high resolution of the PANDA experiment could be particularly advantageous in studying the $\P\*$ states. As near-threshold states, their lineshapes will be intrinsically linked to their underlying meson-baryon degrees of freedom, as in the case of $X(3872)$: see, for example, refs~\cite{Braaten:2007dw,Hanhart:2007yq,Kalashnikova:2009gt,Zhang:2009bv,Danilkin:2010cc,Artoisenet:2010va,Hanhart:2011jz}.

\section{Parallels with X(3872)}
\label{molecule}

There are some intriguing parallels between $\P^*$ and the charmonium-like state $X(3872)$. 

Due to its extreme proximity to threshold, interpretations of $X(3872)$ usually involve $D^{*0}\Db^{0}+D^{0}\Db^{*0}$ degrees of freedom. While models (discussed below) differ on the nature of the binding interaction, a significant meson-antimeson component is unavoidable due to the S-wave coupling and proximity to threshold. It is an interesting exercise to try to find a baryon-antimeson analogue of this meson-antimeson state. Specifically, we search for a pair of charmed baryons $Y_c$ and $Y_c^*$ from which to build a $Y_c^{*+}\Db^{0}+Y_c^+\Db^{*0}$ state. (Note that $Y_c$ and $Y_c^*$ are only being used as labels here: there is no assumption that one is an excitation of the other.) To turn a $D^{(*)}$ meson $c\q$ into a charmed $Y_c^{(*)}$ baryon $cqq$ requires the replacement of one colour antitriplet ($\q$) with another ($qq$). This meson-baryon symmetry has been  widely applied to exotic and non-exotic spectroscopy~\cite{Jaffe:2003sg,Maiani:2004vq,Burns:2004wy,Selem:2006nd,Hernandez:2008ej,Majethiya:2008ia,Eakins:2012fq,Guo:2013xga}.

In the case of $X(3872)$, the thresholds corresponding to the two wavefunction components $D^{*0}\Db^{0}+D^{0}\Db^{*0}$ are obviously degenerate, but this is not true of the analogous $Y_c^{*+}\Db^{0}+Y_c^+\Db^{*0}$ state. In order for there to be strong mixing between the two components, we want their thresholds to be approximately degenerate,
\be
M(Y_c^*)+M(\Db^0)\approx M(Y_c)+M(\Db^{*0}),
\ee
which constrains the masses of the $Y_c\*$ states
\be
M(Y_c^*)-M(Y_c)\approx M(\Db^{*0})- M(\Db^{0})=142.1\pm0.2\MeV .
\label{eq:massdiff1}
\ee
It turns out that there is only one possibility which works well; taking $Y_c$ and $Y_c^*$ as $\Sc$ and $\Lc^{*}$ respectively gives
\be
M(\Lc^{*+})-M(\Sc^+)=139.4\pm0.7 \MeV.
\label{eq:massdiff2}
\ee
We already knew that this combination would work, as we observed (Table~\ref{props}) that the $\LD$ and $\SDh$ thresholds are approximately degenerate. The point is that there are no other obvious combinations $Y_c$ and $Y_c^*$ which will work in this sense. So the most natural analogue of $D^{*}\Db+D\Db^{*}$ is $\Lc^{*}\Db+\Sc\Db^{*}$, which suggests that  $X(3872)$ at $D^{*}\Db/D\Db^{*}$ threshold is likely to be related to $\P^*$ at $\Lc^{*}\Db/\Sc\Db^{*}$ threshold. (The notation $\Lc^{*}\Db+\Sc\Db^{*}$ is not intended to imply that the two components have equal magnitude in the wavefunction, or that their relative phase is positive. Also, charge labels are omitted here and below.)

\begin{figure*}
\vspace{1cm}
\hspace{1cm}
\rput(-0.1,-0.1){$\Db^{0}$}
\rput(-0.1,2.1){$D^{*0}$}
\rput(2.7,-0.1){$\Db^{*0}$}
\rput(2.7,2.1){$D^0$}
\rput(1.6,1.0){$\pi^0$}
\rput(1.3,-0.5){(a)}
~\includegraphics[width=0.13\textwidth]{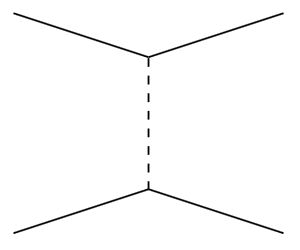}
\hspace{4cm}
\rput(-0.1,-0.1){$\Db^{0}$}
\rput(-0.1,2.1){$\Lc^{*+}$}
\rput(2.7,-0.1){$\Db^{*0}$}
\rput(2.7,2.1){$\Sc^+$}
\rput(1.6,1.0){$\pi^0$}
\rput(1.3,-0.5){(b)}
\includegraphics[width=0.13\textwidth]{pions.png}
\hspace{4cm}
\rput(-0.1,-0.1){$\Db^{*0}$}
\rput(-0.1,2.1){$\Sc^+$}
\rput(2.7,-0.1){$\Db^{*0}$}
\rput(2.7,2.1){$\Sc^+$}
\rput(1.6,1.0){$\pi^0$}
\rput(1.3,-0.5){(c)}
\includegraphics[width=0.13\textwidth]{pions.png}
\vspace{1cm}
\caption{Pion exchange in (a) the  $D^{*}\Db+D\Db^{*}$ state, (b) its $\Lc^{*}\Db+\Sc\Db^{*}$ analogue, and (c) the direct $\SDh$ channel.}
\label{fig}       
\end{figure*}

Many models for $X(3872)$ are based on (or least include) the attractive potential arising from the $D^{*}\Db\to D\Db^{*}$ transition via pion exchange~\cite{Close:2003sg,Tornqvist:2004qy,Swanson:2004pp,Thomas:2008ja,Liu:2008tn,Liu:2008fh,Ding:2009vj,Yu:2011wb,Li:2012cs}. On this basis a molecular state (deuson) at the  $D^{*}\Db$ threshold had been predicted prior to the $X(3872)$ discovery~\cite{Tornqvist:1991ks,Tornqvist:1993ng}. The pion-exchange diagram is shown in Fig.~\ref{fig}(a).

Due to the allowed $\Lc^{*}\to \Sc\pi$ coupling, the $\Lc^{*}\Db\to\Sc\Db^{*}$ transition is also possible through pion-exchange, as depicted in Fig~\ref{fig}(b). It is therefore plausible that pion-exchange in the $\Lc^{*}\Db+\Sc\Db^{*}$ channel is playing a role in $\P^*$. Notice that pion exchange is off-diagonal and mixes the two wavefunction components, just like in the case of $D^{*}\Db+ D\Db^{*}$. In this way the proposed state avoids the argument (Sec.~\ref{models}) that,  because $\Lc\*\to\Lc\*\pi$ is forbidden by isospin,  $\Lc\*\Db\*$  molecules do not form. 

While the analogy between $D^{*}\Db+D\Db^{*}$ and $\Lc^{*}\Db+\Sc\Db^{*}$ is encouraging, there is an important difference. In the former case both channels have a pair of S-wave hadrons, coupled in S-wave. In the latter, one channel is a pair of S-wave hadrons, but the other is an S- and a P-wave hadron. Due to the opposite intrinsic parities in the latter case, it is of course impossible that both channels are coupled in S-wave. The most natural possibility (see Table~\ref{partials}) is Assignment~3, in which the $3/2^-$ state is a mixture of $\SDh$ in S-wave, and $\LD$ in P-wave; other possibilities involve higher partial waves.

If matching $D^{*}\Db+D\Db^{*}$ to $\Lc^{*}\Db+\Sc\Db^{*}$ seems odd, because of the P-wave constituent in one of the channels, the alternative is to disregard the $\LD$ component and consider the analogue of a $D^{*}\Db+D\Db^{*}$ to be a pure $\SDh$ state. (This is similar to the picture of ref.~\cite{Karliner2015}.) Pion exchange in the direct channel $\SDh\to\SDh$ is allowed, as shown in Fig~\ref{fig}(c), so a molecular $\SDh$ state is a possibility. (Even in the mixed state, this direct channel will make some contribution.) In this simpler picture, though, an interesting connection with $X(3872)$ is lost. The kinematics of $\LD\to\SDh$ scattering is very similar to $D^{*}\Db\to D\Db^{*}$ scattering, just because of eqns~\rf{eq:massdiff1} and~\rf{eq:massdiff2}: the pions are just on mass shell. By contrast $\SDh\to\SDh$ scattering is elastic, with off-shell pions. Precisely how the effects of on- and off-shell pions play out in models is a topic of considerable discussion~\cite{Suzuki:2005ha,Braaten:2007ct,Fleming:2007rp,Thomas:2008ja,Baru:2011rs}, but the point remains: from the point of view of kinematics, $\LD\to\SDh$ scattering resembles $D^{*}\Db\to D\Db^{*}$ scattering, while $\SDh\to\SDh$ is rather different.

In any case, the presence of the P-wave $\Lc^*$ constituent may be advantageous from the point of view of binding. Close~\etal.~\cite{Close:2009ag,Close:2010wq} note that for molecules such as $D^{*}\Db+D\Db^{*}$ there is a $q^2$ suppression in the pion-exchange potential for small momentum transfer $q$. This can ultimately be traced to the P-wave couplings associated with each pion vertex $D^*\to D\pi$. They point out that if one of the constituents is replaced by a P-wave meson such as $D_1$, the pion vertices $D_1\to D\pi$ are S-wave. Consequently the potential for molecules with a P-wave meson in each channel, such as $D_1\Db+D\Db_1$, does not suffer threshold suppression and instead behaves as a constant for small $q$. This leads to enhanced attraction and the possibility of deeply bound states. Other authors have considered similar scenarios~\cite{Liu:2007bf,Ding:2008gr,Li:2013bca,Cleven:2013mka,Liu:2013vfa,Wang:2013cya,Chen:2014mwa,Barnes:2014csa}.

Similar effects are to be expected in molecules with meson-baryon constituents. In the $\Lc^{*}\Db+\Sc\Db^{*}$ molecule, one of the vertices ($\Db^*\to\Db\pi$) is P-wave, while the other ($\Lc^*\to\Sc\pi$) is S-wave. In this sense it is intermediary between the $D^{*}\Db+D\Db^{*}$ and $D_1\Db+D\Db_1$ situations, so the prospect of binding seems reasonable. If we assume that the S-wave coupling is beneficial to binding, it is natural that $\P^*$ is more bound with respect to $\Lc^{*}\Db/\Sc\Db^{*}$ thresholds than $X(3872)$ is with respect to $D^{*}\Db/D\Db^{*}$.

While P-wave constituents can benefit from S-wave couplings, they also (on account of those couplings) tend to be rather broad: this can be problematic for the formation of bound states~\cite{Close:2010wq,Filin:2010se}. This should not be a problem in the current case: while $\Lc^*$ benefits from S-wave couplings, it is narrow ($\Gamma=2.6\pm 0.6$ MeV) due to the limited phase space for $\Lc^*\to \Sc\pi$, which follows from eqn~\rf{eq:massdiff2}.

Whether or not pion exchange is sufficient to bind a $\Lc^{*}\Db+\Sc\Db^{*}$ molecule can only be resolved with detailed calculations, beyond the scope of this paper. Recall (Sec.~\ref{decays}) that the total quark spin of $\Lc^*$ (hence also $\LD$) is dominantly $S=1/2$ with a possible $S=3/2$ admixture, while (assuming Assignment~3) $\Sc\Db^{*}$ has $S=3/2$. The pion-exchange transition through the dominant wavefunction component therefore does not conserve spin, but this does not preclude attraction: the analogous potential connecting channels with pairs of S-wave constituents contains a tensor part which mixes states with different quark spin~\cite{Tornqvist:1993ng,Ericson:1993wy}. If there is an  $S=3/2$ component to $\Lc^*$ then spin-conserving pion transitions also contribute.

Another interesting parallel between $X(3872)$ and $\P^*$ is the proximity to three-body threshold. Note that $X(3872)$ couples to $D^0\Db^0\pi^0$ via the decay of either of its heavier constituents $D^{*0}$ or $\Db^{*0}$, and is just above the relevant threshold at $3864.65\pm 0.14$~MeV. Similarly $\P^*$ couples to $\Sc^+\Db^0\pi^0$ via the decay of either of its heavier constituents $\Lc^*$ or $\Db^{*0}$, and in this case the threshold is just slightly above: see Table~\ref{props}. (The physical decay is accessible via the finite width of $\P^*$.) The proximity of three-body thresholds is in a sense not a new aspect of the analogy between $\P^*$ and $X(3872)$, as it follows from eqns~\rf{eq:massdiff1} and~\rf{eq:massdiff2}. It does imply, however, that three-body dynamics for $\P^*$ are likely to be important, just as in the case ~\cite{Fleming:2007rp,Braaten:2007ct,Hanhart:2007yq,Braaten:2007dw,Stapleton:2009ey,Kalashnikova:2009gt,Baru:2011rs} of $X(3872)$.

The analogy between $\P^*$ and $X(3872)$ can be pushed still further. The $X(3872)$ mass coincides not only with the threshold for the open-charm channels $D^{*}\Db/D\Db^{*}$, but also the closed-charm channels $\Jp\omega$ and $\Jp\rho$. Likewise the $\P^*$ mass is not only close to the open-charm thresholds $\Lc^{*}\Db/\Sc\Db^{*}$, but also overlaps exactly with the closed-charm channel $\chip$. It is difficult to ignore this remarkable coincidence.

In some models for $X(3872)$, the proximity of its mass to both open- and closed-charm thresholds is fundamental to its existence. An early molecular model~\cite{Swanson:2003tb} argued that pion-exchange alone is insufficient to bind a molecule, but that the transitions $D^{*}\Db/D\Db^{*}\to \Jp\omega/\Jp\rho$ provide an additional attractive force which leads to binding. Similar results are obtained by Vijande~\etal.~\cite{Vijande:2007fc,FernandezCarames:2009zz,Carames:2012th,Vijande2015}, whose model does not assume pion-exchange, but instead uses the same quark-level interactions as for ordinary mesons and baryons. They find that $D^{*}\Db/D\Db^{*}$ interactions alone are insufficient to bind, but that switching on the $D^{*}\Db/D\Db^{*}\to \Jp\omega$ coupling leads to binding. Similar mechanisms have also been discussed elsewhere~\cite{Dong:2009yp,Danilkin:2011sh,Pena:2012qw}. Note that in the above models, the closed-charm channels are not themselves attractive: it is the coupling between open- and closed-charm which generates attraction. 

It is intuitively obvious that the importance of the $D^{*}\Db/D\Db^{*}\to \Jp\omega/\Jp\rho$ transition is driven by the proximity of the corresponding thresholds, and this was demonstrated explicitly in refs~\cite{Carames:2012th,Vijande2015}. For the same reason, the coupling $\Lc^{*}\Db/\Sc\Db^{*}\to\chip$ is likely to play an important role in $\P^*$. Of course, there are differences, not least that the scattering involves non-zero partial waves. But the enhancement of the scattering due to the small energy denominator is a generic and model-independent effect.

A further similarity between $X(3872)$ and $\P^*$ is isospin violation, already discussed in Sec. \ref{isospin}. In both cases this is due to the mass gap between different charge combinations, namely $D^{*0}\Db^0$ compared to $D^{*+}\Db^-$, and $\Sc^{+}\Db^{*0}$ compared to $\Sc^{++}\Db^{*-}$. A slight difference here is that for $\P^*$ one of the two open-charm channels ($\Lc^{*+}\Db^0$) has only one charge combination and does not contribute to isospin violation.

The parallels between $\P^*$ and $X(3872)$ are compelling, but there is one marked difference which ought to be highlighted. The mass, flavour and $J^{PC}$ quantum numbers of $X(3872)$ allow for a valence charmonium component. In several models $X(3872)$ is essentially a $2\an 3P1$ charmonium distorted strongly by the coupling to $\cc\to D^{*}\Db+D\Db^{*}$ ~\cite{Kalashnikova:2005ui,Li:2009ad,Baru:2010ww,Kalashnikova:2010zz,Hanhart:2011jz,Coito:2012vf,Takizawa:2012hy,Ortega:2012rs,Ferretti:2013faa}, in some cases supplemented by the coupling $D^{*}\Db/D\Db^{*}\to \Jp\omega/\Jp\rho$~\cite{Coito:2010if,Takeuchi:2014rsa}. For $\P^*$ there is no obvious parallel: its quantum numbers allow for mixing with nucleon excitations, but its mass is far above the region of the corresponding states.

This difference notwithstanding, the common features of $X(3872)$ and $\P^*$ are intriguing. In much of the discussion in this paper the $\Lc^{*}\Db$, $\Sc\Db^{*}$ and $\chip$ degrees of freedom were treated as distinct, competing scenarios for $\P^*$. Given the compelling similarities with $X(3872)$, however, it appears more likely that $\P^*$ is a consequence of the interplay among these three degrees of freedom. In this case the picture of distinct, competing scenarios is indeed much too simple; however the conclusions based on that simplified picture are still very useful.

For example, it was argued that  there is the possibility of $I=3/2$ partners for $\P^*$ in the $\SDh$ scenario, but not in the $\LD$ or $\chip$ scenarios (Sec.~\ref{partners}). If all three degrees of freedom are indeed fundamental to the existence of $\P^*$, there cannot be $I=3/2$ partners.

Isospin violation (Sec.~\ref{isospin}) is a feature of $\SDh$  but not $\LD$ or $\chip$ degrees of freedom. If all three degrees of freedom are important to $\P^*$, isospin violation will still be present due to the $\SDh$ component.

The selection rules (Sec.~\ref{decays}) for decays are no longer absolute, as decay channels forbidden for one  wavefunction component are allowed for other components. In principle all of the decays in Table~\ref{dec} (apart from $\Lc\Db\pi$) should be accessible if $\P^*$ contains all three wavefunction components, unless of course there is destructive interference among amplitudes from different components. Experimental observation of $\P^*$ in different modes which are not all allowed by one scenario would be an indication of mixed degrees of freedom.  In this case it may be possible, in principle, to infer the weight of the different components in the wavefunctions from the relative branching fractions in different decay modes; this would of course require detailed and model-dependent calculations.

Relations among decay modes (Sec.~\ref{relations}) would be modified. The $\Jp\D$ and $\eta_c \D$ modes are only accessible through the $\SDh$ component, whereas $\Jp p$ is accessible through all three: in that case eqn~\rf{bf1} no longer applies, but the relative weight of $\Jp\D$ and $\eta_c \D$ is given by eqn~\rf{bf2}. Similarly as $\eta_c p$ is only accessible through the $\LD$ component,  whereas $\Jp p$ is accessible through all three, eqn~\rf{bf3} no longer applies. The $\Sc\Db$ decay would still satisfy eqn~\rf{bf4}, as this is only due to the $I=1/2$ components. On the other hand isospin relations~\rf{bf5} and~\rf{bf6} are modified due to mixing among all three wavefunction components. In the $\Jp N\pi$ mode a mixed state would satisfy neither eqn~\rf{bf7} nor~\rf{bf8}, and this will be a signature of mixing. Finally eqn~\rf{bf9} for $\Lc\Db^*\pi$ would still hold as this is only accessible through the $\SDh$ component.

Most of the proposals for new production modes (Sec.~\ref{production}) would be unchanged, although of course the discovery of the $(I,I_3)=(3/2,-3/2)$ state $\P^{*-}$ would not be expected.

\section{Conclusions}
\label{conclusions}

Due to the proximity of their masses to thresholds, it appears likely that the $\P\*$ states are best described in terms of meson-baryon degrees of freedom. In this paper the model-independent consequences of this assumption have been explored.

Among the various possible $J^P$ assignments,  the experimentally-preferred option (Assignment~1) is also the most natural theoretically, given the relative widths of the states. However, none of the assignments is consistent with naive expectations that threshold effects are expected in S-wave. In kinematic models (which currently match $\P^*$ but not $\P$) this is not particularly problematic, as threshold effects are shown to appear in non-zero partial waves. Molecular models based on pion-exchange can naturally accommodate $\P^*$ in S-wave, and the boson-exchange model can accommodate both $\P$ and $\P^*$. The baryocharmonium model has both states in S-wave, but the connection of their masses to threshold is lost, and deep binding is required.

In all models, the observed charged states would be accompanied by neutral partners, and the mass gaps can indicate the underlying degrees of freedom. In the $\chip$ scenario a gap of around 1 MeV is expected, whereas in the $\LD$ scenario a larger gap of almost 5 MeV is expected.

While several models can accommodate the $J^P$ quantum numbers of the $\P\*$ states, none is yet able to explain why these quantum numbers are unique: in all models several possible $J^P$ are equally plausible, and model calculations are required to determine if partner states are to be expected. In the pion-exchange model, if $\P^*$ is a putative $1/2(3/2^-)$ state, it should be accompanied by a degenerate $3/2(1/2^-)$ partner. Other models predict a richer spectroscopy.

Models based on $\SD$ degrees of freedom are characterised by isospin violation, a necessary consequence of the 5 MeV mass gap separating $\Sc^{(*)+}\Db^{(*)0}$ and $\Sc^{(*)++}\Db^{(*)-}$ thresholds. In this case the observed $\P\*$ states are admixtures of $I=1/2$ and $I=3/2$, and their production and decays will reflect this. An intriguing possibility is that they belong to putative $I=3/2$ multiplets, in which case $(I,I_3)=(3/2,\pm 3/2)$ partners would be expected, although these may not be bound.

Many two- and three-body decay modes are accessible for the $\P\*$ states, and there is an indirect argument suggesting that channels other than the observed $\Jp p$ are significant. The competing scenarios have characteristically different decay patterns, so the experimental observation of $\P\*$ states in different channels can indicate the relevant degrees of freedom. The strongest predictions are due to isospin and the conservation of heavy-quark spin.

In the simple decay model used here, amplitudes for transitions from open- to closed-charm are proportional to the overlaps of quark spin and flavour wavefunctions. In the $\SD$ scenario, significant isospin violation is expected, and due to large wavefunction overlaps the $\Jp \D$ and $\eta_c \D$ modes ought to be comparable to the observed $\Jp p$. In models based on vector meson exchange, these modes are larger still. The $\Sc\*\Db$, $\Jp N\pi$, and $\Lc\Db\*\pi$ modes occur in different charge combinations, and their branching fractions are related by isospin. Experimental measurement of their relative strengths would be a useful indicator of the relevant degrees of freedom. For example, deviation from the $2:1$ ratio for the multi-body $\Jp n\pi^+$ and $\Jp p \pi^0$ channels  indicates isospin violation and the role of $\SD$ degrees of freedom.

Given their discovery in $\Lb\to \Jp p K^-$, the $\P\*$ states may be found in other decay modes in $\Lb\to \P\* K^-$, and in $\Lb\to \Jp p K^{*-}$. The Cabibbo-suppressed mode $\Lb\to\Jp p \pi^-$ has already been observed experimentally, and no $\Jp p$ structures were found: it is important to establish if the absence of a signal is statistically significant, and to compare this to theoretical models. The Cabibbo-suppressed transitions also give unique access to any negatively charged $(I,I_3)=(3/2,-3/2)$ partners to the $\P\*$ states, and relations among different charge modes are indicative of isospin. The lineshapes of $\P\*$ states will be sensitive to their meson-baryon degrees of freedom, and these can be accurately studied at the PANDA experiment.

Finally, there are some intriguing parallels between $\P^*$ and $X(3872)$. Each has mass near to threshold for open-charm pairs ($D^*\Db+D\Db^*$ cf. $\LD+\SDh$), whose channels are coupled by pion-exchange with similar kinematics. Each has mass near to three-body thresholds ($D^0\Db^0\pi^0$ cf. $\Sc^+\Db^0\pi^0$) accessible by the pion decay of its open-charm constituents. Each has mass exactly coinciding with threshold for closed-charm states ($\Jp \rho/\Jp\omega$ cf. $\chip$), so open- to closed-charm mixing should be large. Each has the possibility of isospin violation, arising from the mass gap separating different charge combinations ($D^{*0}\Db^0$ and $D^{*+}\Db^-$, cf. $\Sc^{+}\Db^{*0}$ and $\Sc^{++}\Db^{*-}$). If these similarities are not relevant, they would be remarkable coincidences. An interesting difference between the two is that for the $\P^*$, one of its threshold components includes a P-wave hadron: this implies an S-wave pion coupling with enhanced attraction, consistent with the stronger binding of $\P^*$ compared to $X(3872)$.

\section*{Acknowledgements}

Useful discussions with Qiang Zhao and Jun He are gratefully acknowledged.

\bibliography{//tawe_dfs/users_staff/SFS1/T.Burns/Documents/LaTeX/bibinputs/tjb}

\end{document}